\documentclass[elsarticle,twocolumn,showpacs,superscriptaddress,preprintnumbers,nofootinbib,amsmath,amssymb]{revtex4}

\usepackage{graphicx,color}
\usepackage{dcolumn}
\usepackage{bm}
\usepackage{enumerate}

\usepackage{epstopdf}

\usepackage{amssymb,amsmath,amsthm,mathrsfs}
\usepackage{graphicx}

\begin{document}

\newcommand{\Eq}[1]{Eq. \ref{eqn:#1}}
\newcommand{\Fig}[1]{Fig. \ref{fig:#1}}
\newcommand{\Sec}[1]{Sec. \ref{sec:#1}}

\newcommand{\PHI}{\phi}
\newcommand{\vect}[1]{\mathbf{#1}}
\newcommand{\Del}{\nabla}
\newcommand{\unit}[1]{\mathrm{#1}}
\newcommand{\x}{\vect{x}}
\newcommand{\ScS}{\scriptstyle}
\newcommand{\ScScS}{\scriptscriptstyle}
\newcommand{\xplus}[1]{\vect{x}\!\ScScS{+}\!\ScS\vect{#1}}
\newcommand{\xminus}[1]{\vect{x}\!\ScScS{-}\!\ScS\vect{#1}}
\newcommand{\diff}{\mathrm{d}}

\newcommand{\be}{\begin{equation}}
\newcommand{\ee}{\end{equation}}
\newcommand{\bea}{\begin{eqnarray}}
\newcommand{\eea}{\end{eqnarray}}
\newcommand{\vu}{{\mathbf u}}
\newcommand{\ve}{{\mathbf e}}
\newcommand{\vk}{{\mathbf k}}
\newcommand{\vx}{{\mathbf x}}
\newcommand{\vy}{{\mathbf y}}
\newcommand{\bx}{{\bf x}}
\newcommand{\bk}{{\bf k}}
\newcommand{\br}{{\bf r}}
\newcommand{\bq}{{\bf q}}
\newcommand{\bp}{{\bf p}}
\newcommand{\bv}{{\bf v}}
\newcommand{\TT}{{\rm TT}}
\newcommand{\lqn}[1]{\lefteqn{#1}}
\newcommand{\GW}{{_{\rm GW}}}
\newcommand{\lb}{\left\lbrace}
\newcommand{\rb}{\right\rbrace}
\newcommand{\ab}{{\alpha\beta}}

\newcommand{\uden}{\underset{\widetilde{}}}
\newcommand{\den}{\overset{\widetilde{}}}
\newcommand{\denep}{\underset{\widetilde{}}{\epsilon}}

\newcommand{\dd}{\diff}
\newcommand{\fr}{\frac}
\newcommand{\del}{\partial}
\newcommand{\eps}{\epsilon}
\newcommand\CS{\mathcal{C}}

\def\be{\begin{equation}}
\def\ee{\end{equation}}
\def\ben{\begin{equation*}}
\def\een{\end{equation*}}
\def\bea{\begin{eqnarray}}
\def\eea{\end{eqnarray}}
\def\bal{\begin{align}}
\def\eal{\end{align}}
\def\higgs{h}

\def\TT{{\rm TT}}
\def\GW{{_{\rm GW}}}
\newcommand{\mn}{\mu\nu}
\newcommand{\nn}{\nonumber \\}
\newcommand{\note}[1]{\textcolor{red}{[\textsc{#1}]}}


\title{The Standard Model Higgs as the origin of the hot Big Bang}

\author{Daniel G. Figueroa}
\affiliation{Theoretical Physics Department, CERN, Geneva, Switzerland}
\author{Christian T.~Byrnes}
\affiliation{Astronomy Centre, University of Sussex, Falmer, Brighton BN1 9QH, UK}


\begin{abstract}
If the Standard Model (SM) Higgs is weakly coupled to the inflationary sector, the Higgs is expected to be universally in the form of a condensate towards the end of inflation. The Higgs decays rapidly after inflation -- via non-perturbative effects -- into an out-of-equilibrium distribution of SM species, which thermalize soon afterwards. If the post-inflationary equation of state of the universe is stiff, {\small$w \simeq +1$}, the SM species eventually dominate the total energy budget. This provides a natural origin for the relativistic thermal plasma of SM species, required for the onset of the `hot Big Bang' era. The viability of this scenario requires the inflationary Hubble scale {\small$H_*$} to be lower than the instability scale for Higgs vacuum decay, the Higgs not to generate too large curvature perturbations at cosmological scales, and the SM dominance to occur before Big Bang Nucleosynthesis. We show that successful reheating into the SM can only be obtained in the presence of a non-minimal coupling to gravity {\small$\xi \gtrsim 1$}, with a reheating temperature of {\small$T_{\rm RH} \gtrsim \mathcal{O}(10^{10})\xi^{3/2}(H_*/10^{14}{\rm GeV})^2~{\rm GeV}$}.
\end{abstract}


\maketitle


\section{Introduction}

Compelling evidence supports the idea of an inflationary phase in the early Universe~\cite{Planck2015}. Its specific particle physics realization is however uncertain, so inflation is often parametrised in terms of an inflaton scalar field, with a vacuum-like potential. After inflation, the $reheating$ stage follows, converting all the energy available into different particle species. The latter eventually `thermalize' and dominate the total energy budget; an event that signals the onset of the `hot Big Bang' thermal era.

The details of reheating depend strongly on the model of inflation, and its connection to other matter sectors. Particle production mechanisms in reheating have been investigated in detail, see~\cite{Allahverdi:2010xz,Amin:2014eta} for a review, and references therein. With few exceptions, e.g.~\cite{GarciaBellido:2003wd,Skullerud:2003ki,Bezrukov:2008ut,GarciaBellido:2008ab,Figueroa:2009jw}, most works have focused on understanding the energy transfer from the inflaton into some matter sector, with no connection whatsoever to the Standard Model (SM) of particle physics. However, to reheat the universe successfully, the relativistic thermal plasma dictating the expansion of the universe before Big Bang Nucleosynthesis (BBN), must be dominated by SM species; this is a physical constraint that cannot be evaded. Therefore, even though the inflationary framework is not connected $a~priori$ to the SM, such a connection must exist.

As the Higgs is the only scalar field in the SM, this naturally suggests its role as a gate connecting the SM and inflation. There are essentially three possibilities: 1) the Higgs is identified with the inflaton, 2) the Higgs is not the inflaton but it is coupled to it (either directly or via intermediators), or 3) the Higgs is neither identified with the inflaton nor it is coupled to it. In category 1) we find scenarios where the Higgs gravitational interaction is not minimal, its kinetic term is not canonical, and/or the Higgs is mixed with a hidden sector. Belonging to this category we find different scenarios, e.g.~\emph{Higgs-Inflation}~\cite{Bezrukov:2007ep}, {\it new Higgs-Inflation}~\cite{Germani:2010gm}, {\it Higgs G-inflation}~\cite{Kamada:2010qe,Nakayama:2010sk}, or {\it Higgs-portal} inflation~\cite{Lebedev:2011aq}.

In this paper we will rather consider the inflationary sector characterized, as usual, by a singlet scalar inflaton field {\small$\phi$}, unrelated to the SM Higgs {\small$\mathcal{H}$}. As {\small$|\mathcal{H}|^2$} is the only SM operator of dimension {\small$\Delta = 2$}, Lorentz and gauge invariant, the Higgs bilinear can be coupled to the inflaton, for instance through the scale-free quartic operator {\small$g^2\phi^2|\mathcal{H}|^2$} with dimensionless coupling {\small$g^2$}, or via a trilinear interaction {\small$M\phi|\mathcal{H}|^2$} with {\small$M$} some mass scale\footnote{Couplings of the inflaton with the SM via irrelevant operators of dimension {\small$\Delta > 4$} are also plausible. However the transfer of energy inflaton-to-SM is expected to be more inefficient than in the case of inflaton-Higgs relevant/marginal operators.}. This corresponds to the category 2) above. If we consider e.g.~the scale-free interaction, we learn that in order to avoid spoiling the inflationary predictions, it is required that {\small$g^2 \lesssim g_{\rm max}^2 \sim \mathcal{O}(10^{-3})$} for direct couplings~\cite{Lyth:1998xn}, or even {\small$g^2 \lesssim g_{\rm max}^2 \sim \mathcal{O}(10^{-7})$} for couplings radiatively induced from hidden sectors~\cite{Gross:2015bea}. At the same time, in order to achieve an efficient energy transfer into the Higgs, via non-perturbative broad resonance effects {\it a la preheating}, one needs {\small$g^2 \gtrsim g_{\rm np}^2 \sim \mathcal{O}(10^{-8})$}. The window {\small$g_{\rm np}^2 \lesssim g^2 \lesssim g_{\rm max}^2$} can be therefore rather narrow. Furthermore, the inflaton-induced Higgs effective mass {\small$m_{\mathcal{H}}^2 = g^2\phi^2$} will be sub-Hubble during inflation if {\small$g^2 < g_{\rm min}^2 \sim \mathcal{O}(10^{-10})$}. Therefore, unless the inflaton-Higgs coupling is in the range {\small$g_{\rm min}^2 \lesssim g^2 \lesssim g_{\rm max}^2$}, the Higgs will be a light degree of freedom during inflation. This brings up category 3), where the Higgs is so weakly coupled to the inflationary sector ({\small$g^2 \ll g_{\rm min}^2$}), that in practice it is decoupled from it. We will refer to this condition as the {\it weak coupling limit}. In the case of Higgs-inflaton trilinear interactions or irrelevant operators of dimension {\small$\Delta > 4$}, similar considerations can we put forward, defining the equivalent limit for the corresponding couplings.

It is worth stressing that even though small couplings are typically considered as 'unnatural', that the inflationary constraints discussed above allow only for small coupling strengths. Hence, from this point of view, the weak coupling limit can be simply regarded as a specific choice within the allowed parameter space. We will argue that in this limit, the Higgs is universally excited in the form of a condensate around the time inflation ends. Following inflation, the Higgs condensate decays rapidly into the other SM species, due to non-perturbative parametric effects~\cite{Enqvist:2013kaa, Figueroa:2014aya, KariEtAlNonAbelian14, Figueroa:2015hda, Enqvist:2015nrw}. The SM particles, initially out-of-equilibrium, reach a thermal state soon afterwards. If the equation of state (EoS) of the inflationary sector become sufficiently stiff after inflation, the Higgs and its decay products will eventually dominate the energy budget of the universe; this provides a natural origin for the thermal plasma of SM species needed for the onset of the hot Big Bang thermal era. We will discuss the physical constraints that this mechanism need to satisfy in order to successfully reheat the universe, without spoiling other cosmological observations. 

From now on {\small$m_p = {1\over\sqrt{8\pi G}} \simeq 2.44\cdot10^{18}$ GeV} is the reduced Planck mass, {\small$a(t)$} the scale factor, {\small$t$} is conformal time and a subscript $*$ denotes evaluation at the end of inflation. 

\section{Universal Higgs excitation during inflation, and later decay} 

In the unitary gauge the SM Higgs can be written as a real degree of freedom {\small$\mathcal{H} = \higgs/\sqrt{2}$}, with effective potential {\small$V = \lambda(\higgs)\higgs^4/4$}, where the self-coupling {\small$\lambda(h)$} encapsulates the radiative corrections to the potential~\cite{Casas:1994qy,Casas:1996aq}. Let us characterize inflation as a de Sitter period with constant Hubble rate $H_{*}.$ We require $H_* \gg M_{\rm EW}$, where $M_{\rm EW} \sim \mathcal{O}(10^2)$ GeV is the electroweak scale, in order that the Higgs potential remains quartic. The running of {\small$\lambda$} becomes negative above some critical scale {\small$\mu_c$}, with {\small$\mu_c \sim 10^{11}$ GeV} for the SM best fit parameters~\cite{Degrassi:2012ry,Bezrukov:2012sa,Buttazzo:2013uya}, though this scale can be pushed up to {\small$10^{16}$ GeV}, considering the top quark mass {\small$2-3\sigma$} below its best fit. 

For simplicity we will characterize inflation as a {\it de Sitter} background with physical Hubble rate {\small$H_{*} \leq H_*^{\rm max} \simeq 9\cdot 10^{13}$ GeV}~\cite{Planck2015}. To guarantee the stability of the SM all the way up to inflation, we demand {\small$\lambda > 0$}, considering it as a free parameter, albeit chosen within the reasonable range {\small$10^{-5} < \lambda \lesssim 10^{-2}$}~\cite{Figueroa:2015hda}. Within the weak coupling limit, we can consider two options: 

\begin{enumerate}[(i)]
\item {\it Higgs minimally coupled to gravity\,--}. In this case, the Higgs behaves as a light spectator field during inflation~\cite{DeSimone:2012qr,Enqvist:2013kaa}, performing a random walk at superhorizon scales. In de Sitter space, it reaches an equilibrium distribution within a relaxation time of $1/\sqrt{\lambda}$ efolds, with variance~\cite{Starobinsky:1994bd,Enqvist:2012xn} 
\be\label{eq:VarCaseI}
~~~~~~~\langle \higgs^2 \rangle \simeq {\mathcal{O}(0.1)\over\sqrt{\lambda}}{H_{*}^2}\,.
\ee
In large-field inflation the adiabatic attractor is not reached and this result is corrected \cite{Hardwick:2017fjo}, but we do not expect our results to change significantly. 

\item {\it Higgs non-minimally coupled to gravity\,--}. An interaction {\small$\xi |\Phi|^2 R$} with the Ricci scalar {\small$R$}, is required by the renormalization of the SM in curved space~\cite{Birrell:1982ix,Herranen:2014cua}. If {\small$\xi \lesssim 0.1$}, the Higgs is light and we recover the case {\it i)}. If {\small$\xi \gg 0.1$}, the Higgs is heavy and hence it is not excited during inflation. The sudden drop of {\small$R$} at the transition from the end of inflation to a standard power-law post-inflationary regime, induces however a non-adiabatic excitation of the Higgs, which acquires a variance~\cite{Herranen:2015ima}
\begin{eqnarray}\label{eq:VarCaseII}
~~~~~~~\langle \higgs^2 \rangle \simeq {\mathcal{O}(0.1)\over\sqrt{\xi}}{H_{*}^2}\,.
\end{eqnarray}

\end{enumerate}

In the weak coupling limit the Higgs is therefore always excited in the form of a condensate with a large {\it vacuum expectation value} (VEV): either during inflation [case $i)$] with a typical amplitude {\small$\higgs_{\rm rms} \sim H_{*}/\lambda^{1/4}$}, or around the time when inflation ends [case $ii)$] with typical amplitude {\small$\higgs_{\rm rms} \sim H_{*}/\xi^{1/4}$}. Given the weak dependence, respectively, on {\small$\lambda$} and {\small$\xi$}), the main difference between the two cases, rather than in the amplitude, lies in the scale over which the Higgs condensate amplitude varies: while the correlation length is exponentially large in case $i)$, {\small$H_*l_* \sim \rm{exp} (3.8/\sqrt{ \lambda}) \gg 1$}~\cite{Starobinsky:1994bd}, it is only of the size of the horizon at the end of inflation in case $ii)$, {\small$H_*l_* \lesssim 1$}~\cite{Herranen:2015ima}.

Soon after inflation ends, the Higgs condensate oscillates around the minimum of its potential. Each time the Higgs crosses zero, particle species coupled to the Higgs -- the electroweak gauge bosons and charged fermions of the SM -- are created in non-perturbative bursts 
\cite{Casadio:2007ip,Enqvist:2013kaa,Figueroa:2014aya,KariEtAlNonAbelian14,Figueroa:2015hda,Enqvist:2015nrw}. 
Contrary to the standard case of inflaton preheating, where the inflaton dominates the energy budget of the universe, the Higgs here is rather a sub-dominant energy component of the total budget. One can easily see this by considering the Higgs amplitudes Eqs.~(\ref{eq:VarCaseI}), (\ref{eq:VarCaseII}), from where the ratio of the initial Higgs energy density {\small$\langle V_* \rangle \sim {\lambda\over4}\higgs_{\rm rms}^4$} to that of the inflationary sector {\small$\rho_{\rm Inf} = 3 m_p^2 H_*^2$}, is found as
\begin{eqnarray}\label{eq:InitialRatioEnergies}
r_* \equiv {\langle V_* \rangle \over 3 m_p^2 H_*^2} \sim \delta\times\mathcal{O}(10^{-12})\,\left({H_*\over H_*^{\rm max}}\right)^2~\ll~ 1\,,
\end{eqnarray}
with
\begin{eqnarray}
~~~~~~\delta \equiv 1 \,~{\rm [case}~i){\rm]\,,~~~~~or}~~~~~ \delta \equiv {\lambda/\xi} \,~{\rm [case}~ii){\rm]}
\end{eqnarray}

The post-inflationary decay of the Higgs 
has been studied recently in a series of papers~\cite{Enqvist:2013kaa, Figueroa:2014aya,KariEtAlNonAbelian14,Figueroa:2015hda, Enqvist:2015nrw}. Lattice simulations of the dynamics of the Higgs and the energetically dominant electroweak gauge bosons were carried out in~\cite{Figueroa:2015hda,Enqvist:2015nrw}, incorporating the nonlinear and non-perturbative effects of the SM interactions. During the initial Higgs oscillations, there is an abrupt transfer of energy from the Higgs into the gauge bosons, as expected in broad resonance. Eventually the gauge bosons back-react into the Higgs condensate, and break it apart into higher modes, making the Higgs VEV decrease significantly. The transfer of energy from the Higgs into the SM species ends at a time {\small$t = t_{\rm end}$}, when the (conformal) amplitude of the Higgs condensate stabilizes. This moment signals as well the onset of energy equipartition and a stationary regime, from where the system is expected to evolve towards equilibrium. The time {\small$t_{\rm end}$}, computed within an Abelian approach~\cite{Figueroa:2015hda}, is given by
\begin{eqnarray}\label{eq:tEquiPartition}
t_{\rm end} \simeq 58.9 \beta^{-(1+3w)\over3(1+w)}q_{\rm tot}^{0.42}\,H_*^{-1}\,,~~~~~q_{\rm tot} \equiv {g_Z^2+2g_W^2\over 4\lambda}\,,
\end{eqnarray}
with {\small$g_Z^2,g_W^2$} the {\small$W^\pm,Z$} gauge couplings, {\small$\beta \equiv \sqrt{\lambda}\higgs_*/H_*$} the initial Higgs amplitude, and {\small$w$} the post-inflationary EoS. For reasonable parameter values, {\small$\mathcal{O}(10^2) \lesssim H_*t_{\rm end} \lesssim \mathcal{O}(10^4)$}~\cite{Figueroa:2015hda}, so the Higgs decay after inflation is generically expected to be fast.

The analysis in~\cite{Figueroa:2015hda,Enqvist:2015nrw} describes the dynamics of the Higgs and the dominant decay species {\small$W^\pm, Z$} gauge bosons. The creation of Fermions through parametric non-perturbative effects~\cite{Figueroa:2014aya}, and the decay (scattering) of gauge bosons into (with) fermions, and vice versa, were not included. Therefore, the value of {\small$t_{\rm end}$} given by Eq.~(\ref{eq:tEquiPartition}) should be interpreted only as an indicative scale of the relaxation time of the fields towards equilibrium. In section {\it III.A.3} we will derive a simple estimate of the thermalization time scale, though its precise value will be unimportant in this paper.

\section{Reheating into the SM}

The oscillation-averaged energy density of the Higgs condensate, given the quartic nature of its potential, scales as $1/a^4$ for as long as the Higgs remains homogeneous within its correlation domain. When the Higgs condensate breaks apart into a distribution of other SM species, the energy density of the decay products also scales as $1/a^4$~\cite{Figueroa:2015hda}. Therefore, the energy density of the SM species after inflation scales as relativistic degrees of freedom, {\small$\rho_{\rm SM} = 3m_p^2H_*^2r_*/a^4$}, where we have set $a_*=1$.

The energy density of the inflaton in the period following inflation evolves as {\small$\rho_{\rm Inf} = 3m_p^2H_*^2/a^{3(w+1)}$}, with {\small$w$} the time averaged value of the EoS during that period, dictated by the inflaton potential. The ratio of the energy density of the SM species to the inflaton, evolves as
\begin{eqnarray}\label{eq:EvolutionRatioEnergies}
r(t) \equiv {\rho_{\rm SM}\over \rho_{\rm Inf}} = r_*a^{3w-1} \sim \delta\cdot10^{-12}\left({H_*\over H_*^{\rm max}}\right)^2a^{3w-1}\,,
\end{eqnarray}
with {\small$r_* \ll 1$} [Eq.~(\ref{eq:InitialRatioEnergies})] representing the initial suppression of the energy density of the SM to inflaton. 

The EoS {\small$w$} between the end of inflation and BBN is unconstrained by observations. We require {\small$-1/3 < w \leq 1$} after inflation (by definition {\small$w < -1/3$} during inflation). Although it is typically assumed that {\small$0 \lesssim w \lesssim 1/3$}, there is no reason to exclude a {\it stiff} case {\small$1/3 < w \leq 1$}. This is the case, for instance, of steep inflation~\cite{Copeland:2000hn,Liddle:2003zw} in brane world scenarios\footnote{In these scenarios the Friedmann equation is modified during inflation but it is recovered just after inflation, and then an expansion history with stiff equation of state develops.}. In fact, a post-inflationary stiff EoS can be easily implemented within any inflationary sector. Denoting by {\small$V$} and {\small$K$} the inflaton potential and kinetic energy, during inflation a slow-roll regime {\small$V \gg K$} is typically attained. If a feature in the inflaton potential makes its amplitude drop to {\small$V < K/2$}, this triggers the end of inflation, as the EoS {\small$w = (K-V)/(K+V) > 1/3$} becomes stiff in that moment. The simplest realization of this {\it Kination-domination} (KD) regime~\cite{Spokoiny:1993kt,Joyce:1996cp}, is to assume a rapid transition from {\small$V \gg K$} during inflation, to some small value {\small$V \ll K$} after inflation, the actual value of {\small$V$} being irrelevant. If after inflation, {\small$V = 0$} then {\small$w = +1$}, while if {\small$V/K \ll 1$} but {\small$V \neq 0$}, then {\small$w \simeq +1 - \mathcal{O}(V/K)$}. 

If we define {\small$\delta w \equiv (w - 1/3)$}, then {\small$r(t) = r_*a^{3\delta w}$}. If {\small$\delta w \leq 0$} (the standard assumption of {\small$w \leq 1/3$}), then {\small$r(t)$} either remains as small as {\small$r_*$} ({\small$w = 1/3$}), or decrease even further as {\small$\propto a^{-3|\delta w|}$} ({\small$0 \leq w < 1/3$}). However, for a stiff EoS, {\small$0 < \delta w \leq 2/3$} and {\small$r(t)$} grows. Despite starting from a very small value, {\small$r(t_*) = r_* \ll 1$}, for a stiff EoS there is always a time {\small$t_{\rm SM}$} for which {\small$r(t \geq t_{\rm SM}) \geq 1$}. By construction {\small$1 = r_* a_{\rm SM}^{3\delta w}$}, with {\small$a_{\rm SM} \equiv a(t_{\rm SM}) = r_*^{-1/3\delta_w}$}. Using {\small$a(t) \propto (H_*t)^{2/(2+3\delta w)}$}, we find
\begin{eqnarray}\label{eq:RHtime}
H_*t_{\rm SM} \simeq r_*^{-{(2+3{\delta w})\over 6\delta w}} \sim \left(10^{12}\over\delta\right)^{\hspace*{-1mm}{(2+3{\delta w})\over 6\delta w}}\left({H_*\over H_*^{\rm max}}\right)^{\hspace*{-1mm}-{(2+3{\delta w})\over 3\delta w}}.
\end{eqnarray}
The energy budget of the universe becomes dominated by the SM fields at a time {\small$t = t_{\rm SM}$} after inflation. If the SM particles are already in thermal equilibrium when its dominance begins, one can compute the temperature {\small$T_{\rm SM}$} of the system at {\small$t = t_{\rm SM}$}. Using {\small$\rho_{\rm SM}(t_{\rm SM}) \equiv {\pi^2\over30}g_{\rm SM}T_{\rm SM}^4$} {\small$= 3m_p^2H_*^2r_*/a_{\rm SM}^4$}, it is obtained as
\begin{eqnarray}\label{eq:RHtemp}
T_{\rm SM} 
\simeq {3\over g_{\rm SM}^{1/4}}\cdot 10^{13}10^{-4\over \delta w}\delta^{4+3\delta w\over 12\delta w}\left({H_*\over H_*^{\rm max}}\right)^{{\hspace*{-1mm}(2+3\delta w)\over 3\delta w}}\hspace*{-1mm}{\rm GeV},
\end{eqnarray}
with {\small$g_{\rm SM}$} the SM thermal degrees of freedom at {\small$t_{\rm SM}$}. 

The process just described can be of course identified with a reheating mechanism, identifying  Eq.~(\ref{eq:RHtemp}) with a reheating temperature, but only if certain non-trivial circumstances are met, which we discuss next. 

\subsection{Requirements for successful reheating}

{\it 1) Ensuring small cosmological perturbations\,-}. A sufficiently long period of KD allows the Higgs to generate the total energy density, making the Higgs a curvaton candidate. At {\small$t = t_{\rm SM}$}, the Higgs field perturbations are converted into adiabatic perturbations. In case $i)$, where the Higgs field perturbations were generated during inflation, we have $\delta h\sim H_*$, and the power spectrum generated by the Higgs field using (\ref{eq:VarCaseI}) is $\sim\delta h^2/\langle h^2\rangle\sim\lambda^{1/2}$. Unless $\lambda$ finely tuned to a very small value, the result is far larger than the observed perturbation amplitude of $10^{-9}$ which rules out case $i)$, in agreement with~\cite{Kunimitsu:2012xx}. In case $ii)$ the Higgs is heavy during inflation, leading to its perturbations being exponentially suppressed. Fortunately, this does not lead to a completely smooth universe: unavoidable gravitational couplings between the inflaton and Higgs field mean that the inflaton perturbations are preserved, even after the inflaton energy density becomes negligible at {\small$t > t_{\rm SM}$}~\cite{Sloth:2005yx,Bartolo:2005jg}. Therefore case $ii)$ remains observationally viable, provided that the inflaton field is chosen such that it generates the observed perturbation spectrum.

{\it 2) Ensuring SM dominance before BBN\,-.} For the above mechanism to represent a viable reheating scenario, we need the SM dominance to occur before BBN, i.e.~{\small$T_{\rm SM} > T_{\rm BBN} \simeq$ MeV}. Using Eq.~(\ref{eq:RHtemp}) with {\small$H_* = H_*^{\rm max}$}, and taking {\small$\delta \sim \mathcal{O}(10^{-2})$} ({\small$\delta \sim \mathcal{O}(10^{-4})$}), these conditions imply {\small$\delta w \gtrsim 0.30$} ({\small$\delta w \gtrsim 0.36$}), or equivalently {\small$w \gtrsim 0.63$} ({\small$w \gtrsim 0.69$}). The case when a transition from {\small$V \gg K$} to {\small$V \ll K$} occurs in the inflationary sector, implying {\small$w \simeq +1$}, shows that the `stiffness' requisite is not a strong constraint. Therefore, from now on we will adopt {\small$w = +1$} as a fiducial case. 

It turns out that applying Eq.~(\ref{eq:RHtemp}) for obtaining {\small$T_{\rm SM}$} in the case $ii)$ -- the only viable case -- is misleading. This is because the Higgs field becomes tachyonic once the KD regime is established after inflation, acquiring a mass {\small$m_{\higgs}^2 = -6\xi H_*^2$} ({\small$m_{\higgs}^2 = -3|3w-1|\xi H_*^2$} for arbitrary stiff EoS {\small$1/3 < w \leq 1$}). Eq.~(\ref{eq:RHtemp}) was derived however on the basis of the Higgs amplitudes Eqs.~(\ref{eq:VarCaseI}), (\ref{eq:VarCaseII}), implicitly assuming an outgoing {\small$m^2 \geq 0$} mass state~\cite{Birrell:1982ix}. As the tachyonic condition makes the Higgs amplitude to grow exponentially fast after inflation, {\small$\higgs \propto \exp\lbrace\sqrt{6\xi}\int (\dot a/a)dt\rbrace$}, this actually solves this problem: the Higgs self-interactions will naturally shut-off the tachyonic instability on a time scale much shorter than the initial Hubble time $1/H_*$, when {\small$\lambda \langle\higgs^2\rangle \gtrsim 6\xi H_*^2$} (neglecting the time evolution of the Hubble rate for the simplicity). In order to avoid a cosmological catastrophe with the Higgs reaching a deeper vacuum than the electroweak one~\cite{Espinosa:2007qp,Hook:2014uia,Espinosa:2015qea}, there is a maximum amplitude {\small$\higgs \leq \higgs_{\rm vac}$} that the Higgs should not surpass, with {\small$\higgs_{\rm vac}(m_t,m_H,\alpha_s)$} a function depending sensitively on the top quark mass {\small$m_t$}, the Higgs mass {\small$m_H$}, and the strong coupling constant {\small$\alpha_s$}~\cite{Bezrukov:2012sa,Degrassi:2012ry,Buttazzo:2013uya}. The maximum Hubble rate that maintains vacuum stability is {\small$ H_*^{\rm vac} = \sqrt{\lambda\over 6\xi} \higgs_{\rm vac}$}. The Higgs amplitude and Higgs energy fraction at the moment of tachyonic stabilization are
\begin{eqnarray}\label{eq:VEVafterTachyon}
\langle \higgs^2 \rangle \sim {6\xi\over\lambda}H_*^2 = \higgs_{\rm vac}^2\left({H_*\over H_*^{\rm vac}}\right)^2\,,\hspace*{0.6cm}\\
r\equiv {\langle V\rangle \over 3m_p^2H_*^2}\sim \mathcal{O}(10^{-8}) \times {\xi^2\over\lambda} \left({H_*\over H_*^{\rm max}}\right)^2.\label{eq:FinalRatioEnergies}
\end{eqnarray}

\begin{figure}
\centering
   \includegraphics[width=1\linewidth]{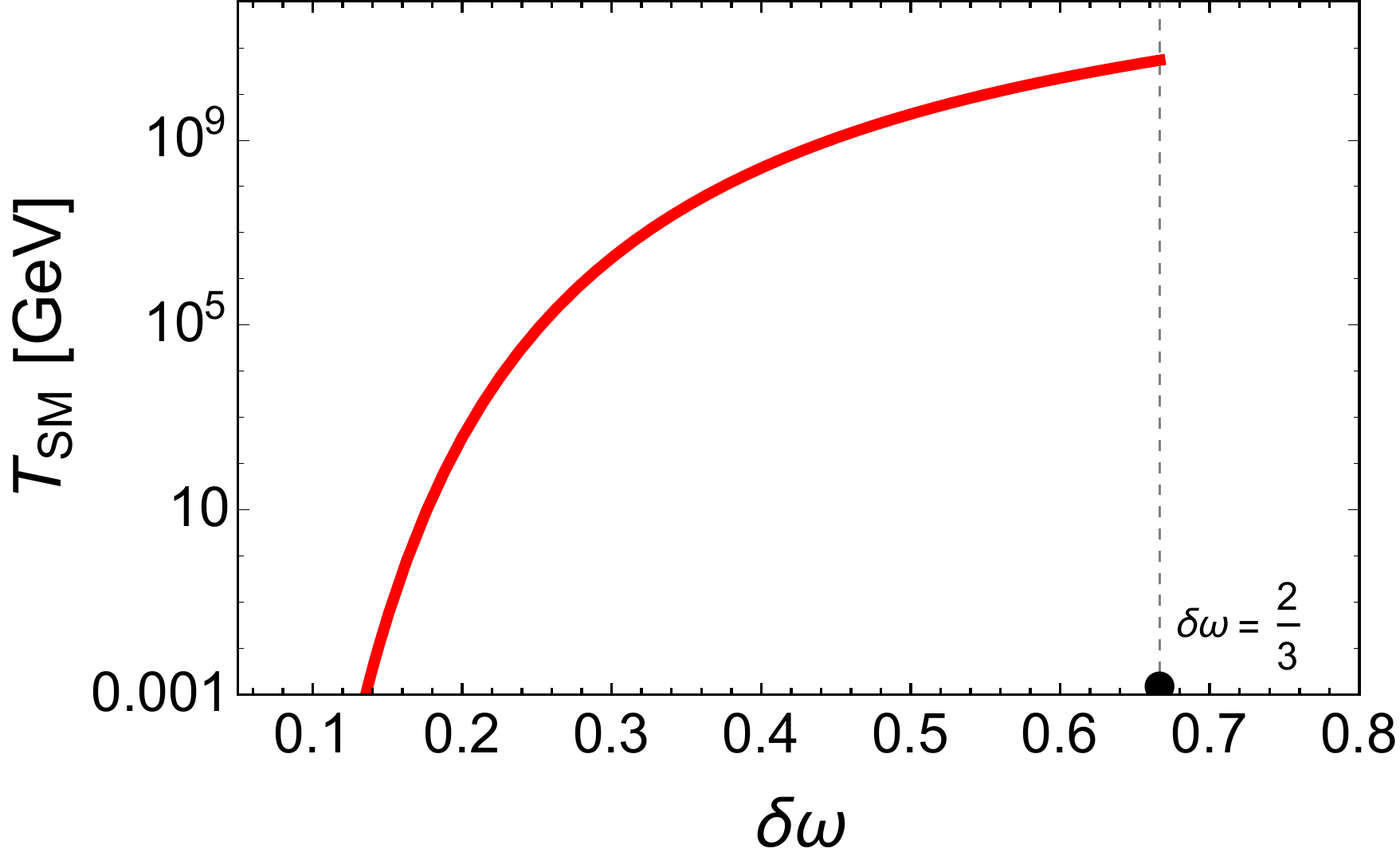}
   \label{fig:Ng1} 
   \includegraphics[width=1\linewidth]{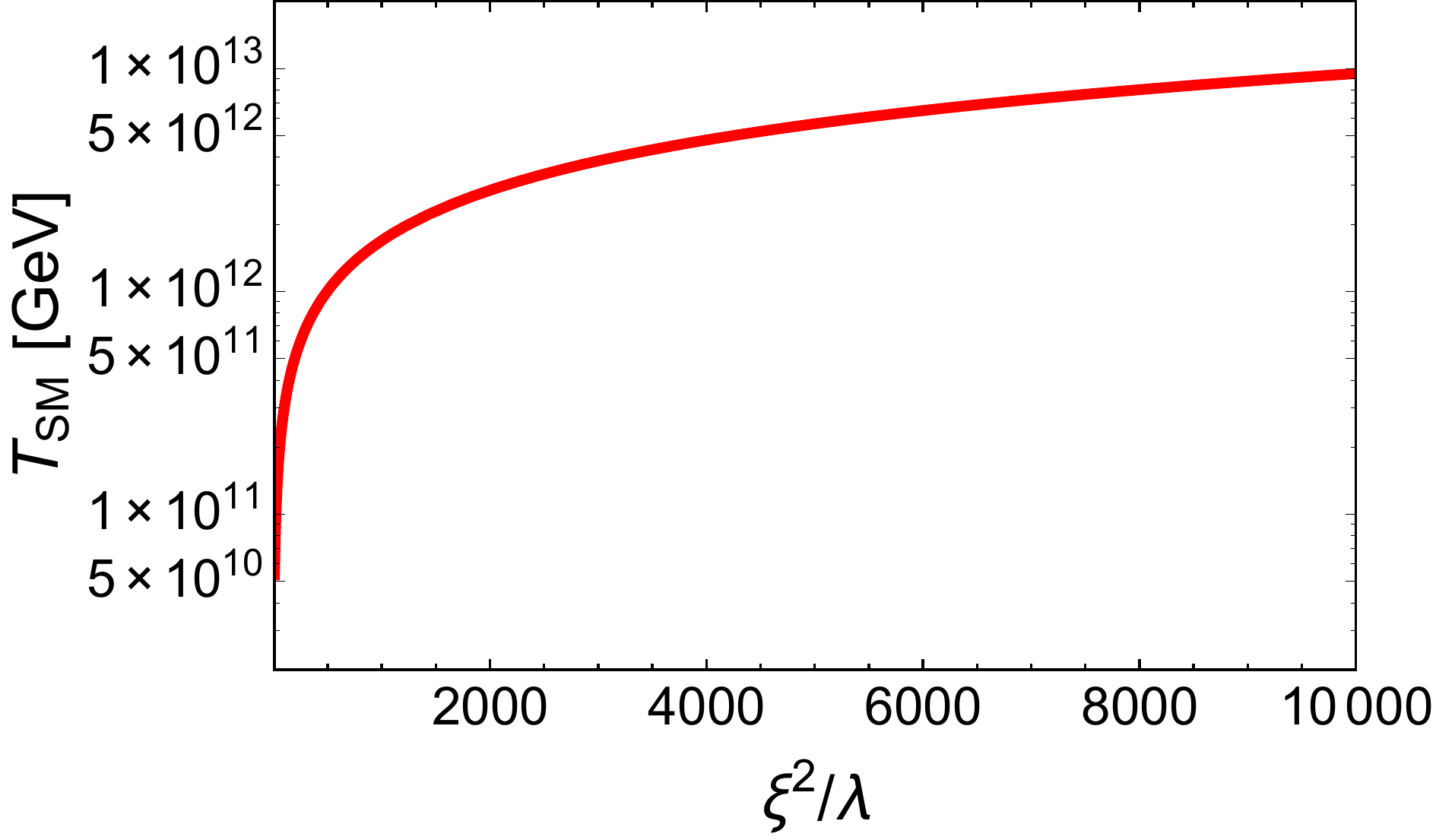}
   \label{fig:Ng2}

\caption[Two numerical solutions]{Upper Figure: Behaviour of Eq.~(\ref{eq:RHtempII}) $T_{\rm SM}$ versus $0< \delta w \leq 2/3$ for fixed values $\xi^2/\lambda = 10$ and $H_*/H_*^{\rm max} = 1$. 
Lower Figure: Behaviour of Eq.~(\ref{eq:RHtempII}) $T_{\rm SM}$ versus $10 \lesssim \xi^2/\lambda \lesssim 10^4$ for for fixed values $\delta w = 2/3$ and $H_*/H_*^{\rm max} = 1$.}
\label{fig}
\end{figure}

The rapid tachyonic phase makes the Higgs amplitude experience a significant growth until the Higgs self-interactions stabilizes the amplitude to Eq.~(\ref{eq:VEVafterTachyon}). Afterwards the non-minimal coupling to gravity quickly becomes unimportant, since $\xi R \sim -6\xi H_*^2/a^6$, so the Higgs oscillates around its potential as if $\xi=0$. To compute the reheating temperature {\small$T_{\rm RH}$} taking into account the impact of the tachyonic phase, we need to use Eq.~(\ref{eq:FinalRatioEnergies}) [instead of Eq.~(\ref{eq:InitialRatioEnergies}), which was used to derive Eq.~(\ref{eq:RHtemp})]. The temperature at the time the SM dominates is
\begin{eqnarray}\label{eq:RHtempII}
T_{\rm SM} &\simeq & {3\over g_{\rm SM}^{1/4}}\cdot 10^{14}10^{-8\over3\delta w}\left(\xi^2\over\lambda\right)^{4+3\delta w\over 12\delta w}\left({H_*\over H_*^{\rm max}}\right)^{\hspace*{-1mm}{2+3\delta w\over 3\delta w}}\hspace*{-1mm}{\rm GeV}\nonumber\\
&\sim& 3\cdot10^{10}\left({\xi^{2}\over\lambda}\right)^{3/4}\left({H_*\over H_*^{\rm max}}\right)^2{\rm GeV}
\end{eqnarray}
where the second line assumes {\small$w = 1 \Leftrightarrow \delta w = 2/3$}. The maximum temperature is obtained using {\small$H_* = H_*^{\rm vac}$}. From {\small$T_{\rm SM} > T_{\rm BBN}\sim 1 MeV$} \cite{Hannestad:2004px}, we deduce that {\small$H_* \geq H_*^{\rm min} \equiv (\xi^2/\lambda)^{-3/8}\cdot 10^{7}$ GeV}. Successful reheating is therefore only possible for relatively large inflationary energy scales. In Fig \ref{fig} we plot the temperature as a function of various model parameters.\footnote{We note the caveat that our calculation does not apply to cases of very low energy-scale inflation, when the Higgs field is not well approximated by a quartic potential.}

Let us note that in reality we need the SM thermal plasma to dominate some time before BBN, so that when BBN is ignited, the expansion rate is sufficiently close to radiation domination. If we demand that $T_{\rm SM} = p T_{\rm BBN}$ with $p > 1$, then the Hubble rate at BBN will be $H(T_{\rm BBN}) = H_{BBN}^{\rm (RD)}\left(1+1/p^2\right)^{1/2}$, where $H_{BBN}^{\rm (RD)}$ is the theoretically correct BBN expansion rate in an exact RD background. The relative difference is then $\left(H(T_{\rm BBN})/H_{BBN}^{\rm (RD)} -1\right)\times 100 \simeq {1\over 2p^2} \%$. Therefore, it is enough that $p \geq 10$ so that the deviation of the initial BBN expansion rate with respect an ideal RD case, is less than $1\%$.

{\it 3) Ensuring thermal equilibrium before SM dominance\,-.} The thermalization time of the SM fields can be estimated as {\small$t_{\rm Eq} \sim 1/(\alpha^2T_{\rm Eq})$}, with {\small$T_{\rm Eq}$} the temperature of the system when thermal equilibrium is first established, 
and {\small$\alpha = g^2/(4\pi)$} the relevant coupling(s). Defining {\small$t_{\rm Eq} \equiv \gamma t_{\rm end}$}, with {\small$t_{\rm end}$} given by Eq.~(\ref{eq:tEquiPartition}), and using {\small$\rho_{\rm Eq} = (g_{\rm Eq}\pi^2/30)T_{\rm Eq}^4 = 3m_p^2H_*^2r_{\rm vac}/a_{\rm Eq}^{4}$}, {\small$T_{\rm Eq} \sim 1/(\alpha^2\gamma t_{\rm end})$} and {\small$a_{\rm Eq} \simeq (2\gamma H_* t_{\rm end})^{1/2}$}, with {\small$g_{\rm Eq}$} the SM thermal degrees of freedom at {\small$t_{\rm Eq}$}, we find {\small$\gamma \sim \mathcal{O}(10^3)/\xi$}, where we have set {\small$g^2 \simeq 0.3$} and {\small$g_{\rm Eq} \simeq 100$}. We confirm $a~posteriori$, therefore, that in fact {\small$t_{\rm Eq} \ll t_{\rm SM}$}. Although a more elaborate calculations of {\small$t_{\rm Eq}$} could be made \cite{Arnold:2002zm,Kurkela:2011ti,Kurkela:2014tea}, the precise value of {\small$t_{\rm Eq}$} is irrelevant for the purpose of reheating the universe into the SM, as long as {\small$t_{\rm Eq} \ll t_{\rm SM}$}. At {\small$t \geq t_{\rm SM}$} the expansion of the universe becomes driven by a thermal relativistic plasma of SM species, as required by the standard hot Big Bang paradigm. The temperature {\small$T_{\rm SM}$} [Eq.~(\ref{eq:RHtempII})] can therefore be identified with a $reheating$ $temperature$, defined as the highest temperature reached by the SM thermal plasma when it first dominates the energy budget. For instance, for KD with {\small$w \simeq 1$} ({\small$\delta w \simeq 2/3$}), {\small$H_* = 0.1H_*^{\rm vac} = 0.01H_*^{\rm max}$}, and {\small$\lambda = 0.005$}, we obtain {\small$T_{\rm SM} \simeq 10^{9}\xi^{3/2}$} GeV.

{\it 4) Other considerations\,-}. The inflaton as well as the Higgs field undergoes a non-adiabatic change in mass during the rapid transition from inflation to KD. The inflaton dominates the energy budget of the universe at this time, 
so even if a small fraction of the inflaton condensate decays into radiation during this transition, then the inflaton decay products might forever dominate over the Higgs energy density, spoiling our goal of achieving the hot Big Bang from the Higgs field. In the limit of a fast transition, the fraction of energy in inflaton decay products immediately after the transition can be estimated as
\begin{eqnarray}
\frac{\rho_{\rm Inf}^{\rm decay}}{3 m_p^2 H_*^2}\sim {m_\phi^4 \over 3 m_p^2 H_*^2} \sim \mathcal{O}(10^{-9}) \times \eta_\phi^2 \left({H_*\over H_*^{\rm max}}\right)^2
\end{eqnarray}
where $m_\phi$ is the effective inflaton mass just before the transition and $\eta_\phi\equiv m_\phi^2/(3H_*^2)<1$ is a slow-roll parameter. Comparing this to the Higgs energy after its tachyonic growth is stabilised, Eq.~(\ref{eq:FinalRatioEnergies}), we see that for $\xi \geq 1$ the Higgs strongly dominates the total radiation component at this time, by a factor $\sim \xi^2/(\lambda\eta_\phi^2) \gg 1$.

\section{Discussion}

The cosmological implications of the SM Higgs in the early Universe remain to be clarified. The possible role of the Higgs as an inflaton or as mediator field connected somehow to the inflationary sector, remains unknown. The circumstances to prevent a catastrophic cosmological instability by which the Higgs might reach a deeper (and negative) vacuum different than the electroweak one during inflation or preheating, has recently triggered a great deal of attention~\cite{Espinosa:2007qp,Enqvist:2013kaa,Figueroa:2014aya,KariEtAlNonAbelian14,Figueroa:2015hda,Enqvist:2015nrw,Enqvist:2014bua,Hook:2014uia,Kobakhidze:2013tn, Kobakhidze:2014xda, Spencer-Smith:2014woa, Shkerin:2015exa, Espinosa:2015qea, Gross:2015bea, Branchina:2015nda, DiVita:2015bha, DiLuzio:2015iua, Grobov:2015ooa, Kamada:2014ufa, Kohri:2016wof, Ema:2016kpf}. 

In this letter we consider the SM stable all the way up to the inflationary scale, and the SM Higgs sufficiently weakly coupled to the inflationary sector; a circumstance that we refer to as the decoupling limit. The Higgs is then universally excited either during or shortly after the end of inflation. We introduce a period of KD with stiff EoS and show that under such circumstance, the Higgs becomes a curvaton which generates unacceptably large perturbations in the absence of a non-minimal coupling to gravity. If a sufficiently large non-minimal coupling to gravity is considered {\small$\xi \gtrsim 1$}, the post-inflationary decay of the Higgs provides a simple explanation for the origin of the relativistic thermal plasma of SM species (the Higgs decay products), necessary to begin the `hot Big Bang' radiation era. Currently the relation between two of the fundamental pillars of our understanding of the Universe, the SM of particle physics and the inflationary framework, is unknown. Therefore, obtaining a mechanism providing an origin of the thermal universe dominated by the SM species is not trivial. The mechanism we propose provides a possible explanation for the reheating of the Universe into the SM fields after inflation, with a reheating temperature that can be rather large. Our major requirement of the inflationary sector is that the background energy density is dominated by the kinetic part after inflation; a condition which is independent of the inflaton potential during inflation. 

A potentially observable consequence of the KD regime after inflation is that the otherwise (almost) scale invariant background of gravitational waves expected from inflation, will be boosted at the high frequency end of the spectrum~\cite{Giovannini:1998bp,Giovannini:1999bh,Boyle:2007zx}. Another consequence, though rather unlikely to be observable, is the production of a background of gravitational waves from the Higgs decay products themselves~\cite{Figueroa:2014aya,Figueroa:2016ojl}, with a peak amplitude today {\small$h^2 \Omega_{\rm GW}^{(o)} \sim 10^{-16}$} (for {\small$H_* = H_*^{\rm max}$}) at {\small$f_p \sim 10^{11} {\rm Hz}$}.

It will be interesting to explore the introduction of an inflation-to-KD transition as  a model-dependent feature in the inflaton potential, as well as a proper study of the thermalization of the SM species after inflation. The need to produce dark matter~\cite{Djouadi:2011aa,Cline:2013fm,Craig:2014lda,Enqvist:2014zqa} and to realize baryogenesis~\cite{Kusenko:2014lra,Yang:2015ida,Pearce:2015nga}, within the setup we are proposing, are also interesting avenues to be explored.

In summary, we have shown for the first time that one can generate the entire post-inflation SM radiation bath from the Higgs field, without spoiling the successful predictions of the observed perturbation spectrum from inflation, and without any contribution or coupling to the inflaton field. We then quantified the required parameter space in which this is possible, and found that an order one or larger non-minimal coupling between the Higgs field and gravity is required in order to not spoil the observed spectrum of primordial perturbations. 

{\it Acknowledgements}. DGF is very grateful to F.~Torrent\'i and J.~Garc\'ia-Bellido for collaboration in a related project, and to L.~Hui, A.~Kurkela, G.~Moore and F.~Piazza, for interesting discussions about this work. CB is supported by a Royal Society Univ.~Research Fellowship.


\bibliography{HiggsReheating}

\end{document}